\documentstyle{article}

\def\be{\begin{equation}}
\def\ee{\end{equation}} \def\eel#1 {\label{#1}\end{equation}}
\def\rz#1 {(\ref{#1}) }

\let\a=\alpha \let\b=\beta \let\g=\gamma \let\d=\delta
\let\e=\varepsilon \let\ep=\epsilon  

   \let\m=\mu
\let\n=\nu    \let\s=\sigma
 \let\o=\omega  
   
\let\O=\Omega

\def\0{\over } \def\1{\vec } \def\2{{1\over 2}} \def\4{{1\over 4}}
\def\5{\bar } \def\6{\partial }

\def\({\left(} \def\){\right)} \def\<{\langle } \def\>{\rangle }
\def\[{\left[} \def\]{\right]}

\def\CX{{\cal X}} \def\CS{{\cal S}} \def\CA={{\cal A}}

\begin{document}

\begin{titlepage}
\renewcommand{\thefootnote}{\fnsymbol{footnote}}
\renewcommand{\baselinestretch}{1.3}
\hfill  TUW - 94 - 21 \\
\vbox{\hfill PITHA - 94 - 49} \\
\vbox{\hfill hep-th/9411163} \\
\medskip
\vfill

\begin{center}
{\LARGE {Poisson-$\sigma$-models:\\
A generalization of 2d Gravity-Yang-Mills systems}
 \\ \medskip  {}}
\medskip
\vfill

\renewcommand{\baselinestretch}{1} {\large {PETER
SCHALLER\footnote{e-mail: schaller@email.tuwien.ac.at} \\
\medskip \medskip
Institut f\"ur Theoretische Physik \\
Technische Universit\"at Wien\\
Wiedner Hauptstr. 8-10, A-1040 Vienna\\
Austria\\
\medskip\medskip\medskip\medskip
THOMAS STROBL\footnote{e-mail:
tstrobl@thphys.physik.rwth-aachen.de} \\ \medskip\medskip
Institut f\"ur Theoretische Physik  \\
 RWTH Aachen\\ Sommerfeldstr. 14, D 52056 Aachen,\\ Germany
} }
\end{center}

\vfill
\begin{center}
Talk delivered at the Conference on Integrable
Systems, \\
JINR, Dubna, 18.-21.7.94
\end{center}
\vfill
\renewcommand{\baselinestretch}{1}                          

\begin{abstract}
A new class of two dimensional integrable
field theories, based on the mathematical notion of Poisson
manifolds, and containing gravity-Yang-Mills systems as well as the G/G
gauged Wess-Zumino Witten-model, are presented. The local
solutions of the classical equations of motions as well as a scheme for
the quantization in a Hamiltonian formulation is
presented for the general model. Partial results of a calculation of
the partition function
on arbitrary Riemann surfaces via path integral techniques are given.
\end{abstract}

\vfill
\hfill  October 1994  \\
\end{titlepage}

\section{Introduction}

In recent years two dimensional topological and almost topological
field
theories, i.e theories involving constraints whose implementation
leaves a finite number of degrees of freedom only, have become an
active field of research \cite{Blau}.
In particular pure gravity and non abelian
gauge theories or the G/G gauged Wess-Zumino-Witten (GWZW) model have
been studied extensively \cite{CGHS},\cite{gravmany},\cite{Rajetal},
\cite{GWZW}.
Striking similarities occurring in the
analysis of these theories strongly suggest that there is a common
mathematical structure behind them.

The aim of this talk is to show that this common mathematical structure
actually exists and is the one of Poisson manifolds: To any Poisson
structure on any finite dimensional manifold one can
associate a two dimensional topological field theory
\cite{LNP},\cite{p9}. Further
nontopological theories are obtained by adding a nontrivial
Hamiltonian term to the action. Pure gravity and non abelian gauge
theories as well as the GWZW model are special examples for the
general class of models (Poisson-$\sigma$-models) formulated in this
way and the formalism allows for their comprehensive treatment.  Under
some natural restrictions on the Poisson manifold serving as target
space and up to
complications arising from a nontrivial world sheet topology the
Poisson-$\sigma$-models are completely integrable.

The talk is organized as follows: After shortly outlining some
features of Poisson manifolds, the general model will be presented.
Examples will be given. In particular the different theories mentioned
above will arise by specific choices of the Poisson structure
(section 2). The integrability of the general model will be outlined
and the local solutions of the classical equations of motion will be
given. A scheme for the quantization in a Hamiltonian formulation
(effectively restricting the topology of the world sheet to the one of
a cylinder) will be presented. This scheme relates the quantization of
the Poisson-$\sigma$-model to the quantization of finite dimensional
systems defined by the Poisson structure on the target space (section
3). We will close the talk by an outlook to open problems. This
outlook will include some remarks on the calculation of the partition
function for the world sheet being an arbitrary genus g Riemann
surface (section 4).

\section{The general model}

Denote by $N$ a manifold of arbitrary dimension with a
Poisson structure. The latter may be expressed either via
Poisson brackets or in terms of an antisymmetric tensor field $P\in
\wedge^2TN$. In local coordinates $X^i$ on $N$ it reads
\be
 P={1\over 2} P^{ij}(X)\6_i\wedge\6_j \quad \sim \quad
  \{ f(X),g(X)\} =P^{ij} f,_i g,_j
\eel Poiss
and is subject to the Jacobi identity
\be
 P^{i[j} P^{lk]}{}_{,i} =0 \, \,  \iff \,\,
 \{ f,\{ g,h\}\} + cycl. =0 \, .
\eel Jaco
In the above equations we used the abbreviations
\be
 \6_i={\6 \over \6X^i} \, ,\quad f,_k =\6_k f \, .
\eel abbr
Summation over double indices is assumed and the square brackets in \rz
Jaco denotes antisymmetrization.

In contrast to the Poisson bracket on the phase space of a Hamiltonian
system, $P$ is not required to be nondegenerate. So in general $P$ will
not induce a symplectic form on $N$. In this case the map $T^ *N \to
TN$, mapping a one-form $\alpha_i dX^i$ to its contraction
$\alpha_i P^{ij}\6_j$ with the tensor field $P$, is not
surjective. As a consequence of the Jacobi identity, however, the
image of this map forms an involutive system of vectorfields and (by
the Frobenius theorem) defines an integral surface $S$ (symplectic
leaf) through any point $X_0 \in N$. From this construction it is
obvious that $P$ can be restricted to yield a nondegenerate
Poisson structure on $S$. The inverse of this Poisson structure
is a symplectic form $\Omega_S$ on $S$, justifying the notion of
symplectic leaves.

There is an alternative description of this construction: If $P$ is
degenerate, one may find functions $f$ on $N$, whose Hamiltonian
vectorfields $X_f = f_{,i}P^{ij}\6_j$ vanish (implying $\{f,.\}=0$).
We will denote these functions as Casimir functions. Let $\{C^I\}$ be a
maximal set of independent Casimir functions. Then $C^I(X)=const
=C^I(X_0)$ defines a level surface through $X_0$. In many interesting
examples the symplectic leaves of $P$ can be identified with
(the connected components of) such level surfaces. In these cases we
will call $P$ integrable.

A simple illustrative example is provided by the Poisson structure
$\{X^i,X^j\}=\varepsilon_{ijk}X^k$ on $R^3$. Obviously $r^2 = X^iX^i$
is a Casimir function and the symplectic leaves are the two-spheres
with constant radius.

As this is a conference on integrable
systems, we will assume in the rest of this talk that $P$ is
integrable. (Many of the following considerations, however, can
be generalized easily to the case where $P$ is not integrable).

Due to the Darboux theorem the symplectic form $\Omega_S$
can be given the coordinate expression $\Omega_S=
dX^1\wedge dX^2 + dX^3 \wedge dX^4 + ...$ by a special choice of local
coordinates $X^\alpha$ on $S$. Together with the Casimir functions
this gives rise to a coordinate system $\{X^I,X^\alpha\}$ on $N$ with
$P^{IJ} = P^{I\alpha} =0$ and $P^{\alpha\beta}=const$. A set of
coordinates with this property will be denoted as Casimir-Darboux
coordinates in the following.

Now denote by $M$ the two dimensional world sheet of a field theory.
As dynamical variables we take the $X^i$ and a field $A$
which is one-form
valued on $M$ as well as on $N$. With coordinates $x^\mu$ on $M$, $A$
can be written as $A= A_{\m i}dx^\m \wedge dX^i$.
In these coordinates the action of the topological
Poisson-$\sigma$-model is given by
\be
  L_{top} = \int_M dx^\m \wedge dx^\n \, [A_{\m i}(x) X^i{}_{,\n}(x) +
  \2 P^{ij}(X(x)) A_{\m i}(x) A_{\n j}(x)]  \, .
\eel actioncoor

With the shorthand notation $ A_i(x)=A_{\mu i}(x)dx^\mu$ it takes the
form
\be
  L_{top}=\int_{M}A_i\wedge dX^i+{1\over 2} P^{ij}A_i\wedge A_j \, .
\eel acti2
As the integrand is a two form on $M$, we can integrate it without any
reference to a metric on $M$. This establishes the topological nature
of the model.

The symmetries of \rz acti2 are induced by the Hamiltonian
vectorfields on $N$
\be
 \d_\ep X^i =  \ep_i(x) P^{ij} \, , \quad
\d_\ep A_i = d\ep_i + {P^{lm}}_{,i} A_l \ep_m \, .
\eel symme
One should note here that the action is invariant under
arbitrary diffeomorphisms on $N$, if $P^{ij}$ is transformed properly.
As $P$ is not a dynamical field, however, gauge transformations
are induced by
diffeomorphisms in the direction of Hamiltonian vectorfields only, as
these diffeomorphisms leave $P$ invariant.

By construction the action  $L_{top}$ is also
invariant under diffeomorphisms of $M$. So one could wonder,
how this invariance is encoded in (\ref{symme}). We will come back to
this question in a minute.

With the additional input of a volume form $\e$ on
$M$ and the choice
of a Casimir function $C$ we can extend the action by adding
\be L_C= \int \e  C(X(x)) \, \eel S+
to $L_{top}$ without spoiling the symmetries (\ref{symme}).
We loose, however, the topological nature of the model.

{}From the action
\be
  L=L_{top} + L_C
\eel gesact
 the equations of motion follow immediately:
 \be
 \begin{array}{c}
  dX^i +  P^{ij} A_j=0 \, , \\[2pt]
  dA_i + \2 P^{lm}{}_{,i} A_l \wedge A_m + C_{,i}=0 \, .
  \end{array}
\eel equom

With \rz equom it is easy to answer the above question about the
invariance
of $L_{top}$ under diffeomorphisms on the world sheet:
For $C=0$ a
diffeomorphism in the direction of a vectorfield $\xi\in TM$ is
generated on-shell by the field dependent choice
$\ep_i=\xi^\m A_{\m i}$ in (\ref{symme}).

Let us now turn to special cases of the model, proving its
universality. Firstly let us consider the case of a nondegenerate
Poisson structure. $P$ then  has  an inverse  $\O$, which is a
symplectic two-form on $N$. Since the action \rz acti2 is quadratic in
the fields $A_i$, we may integrate them out by means of their equations
of motion. Solving the first equation \rz equom for the $A_i$
and plugging the result  $A_i = \O_{ji} dX^j$ back into the action,
we obtain
\be L_{top} \leadsto \int_M \O_{ij} dX^i \wedge dX^j \, .\eel path
This Lagrangian is the starting point for Witten's topological
$\s$-model \cite{WittenSigma} in the Baulieu-Singer approach
\cite{Bauli}.

Next we turn to linear Poisson structures on a linear space $N$. I.e.\
$P^{ij}= f^{ij}{}_k X^k$ or, equivalently,
\be
\{ X^i,X^j\} = f^{ij}{}_k X^k \, .
\eel lie
Due to the Jacobi identity \rz Jaco the constant coefficients
$f^{ij}{}_k$
have to be structure constants of some Lie algebra $h$, and
we may identify $N$ with  $h^*$, the dual of  $h$. The
topological part \rz acti2 of our action is now seen to take the form
of a $BF$-gauge theory:
\be L_{top} = \int_M X^i F_i , \ee
where $F_i \equiv dA_i + \2  f^{lm}{}_i A_l A_m$ is the standard
curvature two-form of the gauge group corresponding to $h$.
Choosing, furthermore, $C$ to equal the quadratic
Casimir of $h$, \rz gesact may be seen to provide nothing but
the first order formulation of a 2d Yang-Mills theory
\be L = \int_M tr \, F \wedge \ast F  \, . \ee
The quantization of this action, performed first in \cite{Migdal},
has experienced much renewed interest within  recent years (cf., e.g.,
\cite{Rajetal}).

A somewhat more complicated situation arises  when choosing a
nonlinear Poisson tensor $P$ on a linear space $N$. It is interesting
to observe that the corresponding topological field theories
incorporate models allowing for a gravitational interpretation. This
may be realized as follows:
Choose $N$ to be three dimensional. Then $P^{ij}= \e^{ijk} u_k(X)$.
The  choice $u_a=\eta_{ab} X^b$, $u_3=V(X^a \eta_{ab} X^b, X^3)$, where
$a, \, b \in \{ 1,2 \}$ and $\eta_{ab} \equiv diag (1,\pm 1)$, solves
the Jacobi identity (\ref{Jaco}) for arbitrary $V$. Next we interpret
the first two one-forms $A_a$ as the {\em zweibein} $e_a$  and the
third one, $A_3$, as the  spin connection $\o$ of a two dimensional
gravity theory. $L_{top}$ may then be rewritten  as
\be L_{top}=\int_M  X_a De^a + X^3 d \o + \2 V  \e_{ab} e^a \wedge e^b
\, , \eel gravaction
where the indices have been raised and lowered with the frame metric
$\eta_{ab}$ and $De^a \equiv de^a + \e^a{}_b \o e^b$.

We want to argue that this action may be regarded as a model for pure
gravity in two dimensions.  Firstly it is invariant against
diffeomorphisms and local frame rotations (of Minkowskian or Euclidean
type as corresponds to the respective choice for $\eta$). Secondly it
is formulated in terms of {\em zweibein} and spin connection.  The
fields $X^i$ might be interpreted as dilaton-like auxiliary fields.
Equivalently their role as generalized momenta for the gravity
variables in a Hamiltonian framework (cf.\ next section) might be
stressed. Thirdly the study of the corresponding classical solutions
leads to a variety of interesting Penrose diagrams including, e.g.,
those of Schwarzschild or Reissner-Nordstroem \cite{Kloesch}. And,
last but not least, \rz gravaction may be seen to yield already
accepted models of 2D gravity for some specific choices of the
potential $V$.  (Note that the 2D analogue of the Einstein-Hilbert
action is basically empty, since it is  a total divergence; so there
is a need for alternative Lagrangians).

Let us illustrate the last item by means of an example,
namely $V \propto (X^3)^2 + const$.
Integrating out the variable $X^3$, which now enters quadratically
into (\ref{gravaction}), we get a term proportional to
$d \o \wedge \ast d \omega$, where $*$ denotes the  operation of
taking the Hodge dual. Now $\e^a{}_b d \o$
is nothing but the curvature two-form of the gravity theory.
$De^a$, on the other hand, is its torsion two-form. Integrating out
$X^a$ we thus get the zero torsion condition, which uniquely
determines the spin connection in terms of the metric.
After the elimination of the $X^i$ the
resulting Lagrangian  takes the form
\be L_{top} \leadsto \int_M d^2x \sqrt{\pm g} (R^2 + const) \, , \ee
where $R=2 \ast d\o(g_{\m\n})$ is the standard Ricci scalar
and $g$ denotes
the determinant of the metric $g_{\m \n}$. So our action becomes the
one of $R^2$-gravity with cosmological constant. Also many other
popular gravitational models may be found to result from
\rz gravaction by appropriate
choices for $V$ (cf., e.g., \cite{Kunst}).
Much of the interest in such models stems from the fact that they
provide a scenario for tackling conceptual issues of quantum gravity
such as the 'problem of time' \cite{CQG}, \cite{LNP} or black hole
radiation \cite{CGHS}.

As a last example let us choose $N$ to be some semisimple group
manifold $G$ and $P$ to be a Lie Poisson structure on it
\cite{Semenov}. It is the subject of \cite{Anton} to show that in this
case the action \rz acti2 is equivalent to the action of the
completely gauged Wess-Zumino-Witten model,
 \be
  \begin{array}{r@{}l}
    GWZW&(g,a_+,a_-)= \2 \int  tr \, (\6_\mu gg^{-1}
          \6^\mu gg^{-1})\, d^2x +
          {1 \0 3} \int\, tr d^{-1}(dgg^{-1})^3 \\[2pt]
                   &+\int tr (a_+\6_-g g^{-1}-a_- g^{-1}
                      \6_+g -a_+ga_- g^{-1}
                     +a_+a_- ) \, d^2x \, .
  \end{array}
 \eel agwzw
In \rz agwzw $g \in G$ plays the role of the target space coordinates
$X^i$. The relation of $a$ and $A$ is not that straightforward and we
want to refer here to \cite{Anton}. Obviously the topological nature of
the GWZW model is much more transparent in the form \rz acti2 than in
its original formulation (\ref{agwzw}).

\section{Classical solutions and quantum states}

Choosing Casimir-Darboux coordinates $\{X^I,X^\a\}$ on $N$, the
Casimir function $C$ in the action \rz gesact is a function of the
$X^I$ only  and
the equations of motion \rz equom simplify to
\be
 dX^I=0 \, , \quad
 dA_I = -C_{,I} \, , \quad
 A_\alpha = (\O_S)_{\beta\alpha} dX^\beta \, .
\eel eomcd
So the
$X^I(x)$ have to be constant on $M$, but otherwise arbitrary, whereas
the $X^\alpha(x)$ remain completely undetermined by the field
equations.
Any choice of the latter determines $A_\alpha$ uniquely through
the last equation in (\ref{eomcd}).
Each of the $A_I$ is determined up
to an exact one-form only.

Still one has not made use of the gauge freedom. As is obvious from
\rz symme any choice of the $X^\alpha$ is gauge equivalent,
and also $A_I \sim A_I
+ dh_I$, where the $h_I$ are arbitrary functions.
Thus locally any solution to the field equations is uniquely
determined by the constant values of the $X^I$.

Additional structures evolve, if global aspects are
taken into account. These will be dealt with elsewhere.

For a Hamiltonian formulation of the theory let us assume $M$ to be of
the form $S^1\times R$ parameterized by a 2$\pi$-periodic coordinate
$x^1$ and the evolution parameter $x^0$.
As the action \rz gesact is already in first order form, the
derivation of the corresponding Hamiltonian system is simple. The zero
components $A_{0i}$ of the $A_i$  play the role of Lagrange
multipliers giving rise to the system of first class constraints
($\6=\6 /\6 x^1$)
\be
 G^i\equiv \6X^i +  P^{ij} A_{1j} \approx 0 \, .
\eel const
The fundamental Poisson brackets (not to be confused with the
Poisson brackets on the target space $N$) are given by
\be
  \{ X^i(x^1), A_{1j}(y^1) \} = - \d^i_j \d(x^1-y^1)
\eel poibr
and the Hamiltonian reads
\be
  H= \int dx^1 (C  - A_{0i}G^i) \, .
\eel hamil

So to any point $x^1\in S^1$ and any index $i$ there corresponds
one dynamical degree of
freedom (represented by $X^i(x^1)$ and the conjugate momentum
$A_{1i}(x^1)$) and one
constraint in (\ref{const}). Thus, by naive counting, one could think
that implementation of the constraints will not leave any degree of
freedom. This is in contradiction to our result for the local
solutions of the equations of motion, characterized by the choice of
constant values for the $X^I$. Indeed, a careful analysis shows that
the constraints are not completely independent. This is most easily
seen in Casimir Darboux coordinates: The components of the
constraints corresponding to the Casimir indices are total
derivatives ($G^I=\6X^I$) and thus
$\oint G^I dx^1 =0$. In correspondence to this dependence of the
constraints, the zero modes $X_{(0)}^I = \int_{S^1} X^I$
of the $X^I$ are neither determined by the constraints, nor are they
affected by gauge transformations.

It is interesting to note that the $X_{(0)}^I$ and the $G^i$ form a set
of conserved quantities covering half of the coordinates in the
unconstrained phase space. Conserved here means that they commute with
the Hamiltonian \rz hamil for any choice of the Lagrange multipliers
$A_{0i}$. Furthermore they commute among each other on the constraint
surface, which is invariant under the flow of the Hamiltonian. So
the constraint
surface forms an integrable system with
respect to the dynamics on the unconstrained phase space.
This explains, why we could find the local solutions of the equations
of motion quite easily. (From this point of view the transition to
Casimir-Darboux coordinates
in \rz eomcd may be seen as a sort of Baecklund transformation,
linearizing the equations of motion).

To quantize the system in an $X$-representation we consider quantum
wave functions as complex valued functionals on the space $\Gamma_N$
of parameterized smooth loops in $N$:
\be
 \Gamma_N=\{ \CX : S^1\to N,x\to X(x) \} \, .
\eel loops
Following the Dirac procedure, only such quantum states are
admissible which satisfy the quantum constraints
\be
 \hat G^i(x) \Psi[\CX ] = \left(
 \6 X^i(x) + i\hbar P^{ij}(X) {\d \0 \d X^j(x)} \right)
 \Psi[\CX ] =0 \, .
\eel quanc

To find the solution of (\ref{quanc}), let us again start with
Casimir-Darboux
coordinates: Then the  $I$-components of the constraints, $\partial
X^I\,\Psi(\CX )$, are seen to  restrict
the support of $\Psi$ to loops which are
contained entirely in some symplectic leaf $S$.

Let us denote the
space of symplectic leaves by $\cal S$.
For $S\in \cal S$ let $\Gamma_S$ be the space of loops on
$S$.
The ansatz $\Psi\vert_{\Gamma_S}=\exp\Phi$ for the restriction of
$\Psi$
to $\Gamma_S$ allows to rewrite the
remaining constraint equations according to
\be
 {\cal A}=-i\hbar \d\Phi \, ,
\eel horco
where $\d$ denotes the exterior derivative on $\Gamma_S$
and $\cal A$ is the one-form on $\Gamma_S$ given by
\be
 {\cal A} = \oint_{S^1} \d X^\a(x)({\O_S})_{\a\b}(X(x))
  \6 X^\b(x)dx \, .
\eel u1con
(Reinterpreting $\cal A$ as a connection in a $U(1)$-bundle over
$\Gamma_S$ the constraint may be seen as a horizontality condition on
the section  $\Psi\vert_{\Gamma_S}$).

Equation \rz horco seems to suggest that $\cal A$ has to be exact.
This is not
true, however, as $\Phi$ is determined up to an integer multiple of
2$\pi$ only. Therefore, $\Psi$ is well defined, iff ${\cal A}$ is
closed
and integral, i.e.\ the intgral of ${\cal A}$ over any closed loop in
$\Gamma_S$ is an integer multiple of $2\pi \hbar$.
To reformulate \rz u1con in a coordinate independent way consider a
path $\gamma$ in $\Gamma_S$. As $\gamma$ corresponds to a one
parameter family of loops in $S$, it spans a two dimensional surface
$\sigma (\gamma )$. $\cal A$ may now be defined alternatively via
\be
 \int_\gamma {\cal A} = \int_{\s(\g)}\Omega_S \, . \label{altda}
\ee
As any closed loop in $\Gamma_S$ generates a closed surface in $S$,
$\cal A$ is closed and integral, iff $\Omega_S$ is closed and integral.
The first condition holds, as $\Omega_S$ is symplectic. The integrality
condition, however, may yield a restriction of the support of $\Psi$
to loops over elements of a (possibly discrete) subset of $\cal S$,
if the second homology
of the symplectic leaves is nontrivial. For $S$ in this subset, $\Psi$
is determined up to a multiplicative constant on any connected
component of $\Gamma_S$.
As the space of connected components of $\Gamma_S$ is in one to one
correspondence with the first homotopy group of $S$, we may identify
physical states with complex valued functions on $\cal I$ defined via
\be
 {\cal I}=\bigcup_{S\in \tilde \CS} \Pi_1(S) \qquad \, \, ,\quad
 \tilde \CS =\{ S\in \CS : \O_S\,
  \hbox{integral}\} \, .
\eel sphst
If the first homotopy of the symplectic leaves is nontrivial, the
quantum states will depend on discrete quantum numbers parameterizing
$\Pi_1(S)$. If the second homology of the symplectic leaves is
nontrivial also (some of) the coordinates parameterizing $\tilde \CS$
may be discrete.

The Hamiltonian (\ref{hamil}), being constant on each of the symplectic
leaves, defines a function on $\cal I$ and thus becomes a
multiplicative operator upon quantization.
Obviously these results are coordinate independent, although we
intermediately used Casimir-Darboux coordinates to derive them.

The above description of physical states is rather formal. Let us
supplement it
by giving an explicit expression for the wave functions:
\be
   \Psi (X(x))=\hat \Psi (X^I,m,n) \, \, \exp i\hbar
\int_{X(x)}d^{-1} \Omega_S \, .
\eel wavfu
Here the $X^I$ are restricted to (possibly discrete) values
characterizing integral symplectic leaves and the integers $m$ and $n$
are understood to parameterize the first and second homotopy of the
corresponding level surfaces. $\hat \Psi$ is an arbitrary function of
these quantum numbers. The notion $d^{-1}\Omega_S$ is formal,
of course. For integral symplectic leaves, however, the
exponential factor in \rz wavfu is well defined via $\int_{X(x)}
d^{-1}\Omega_S = \int_\sigma \Omega_S$ with $\6\sigma = X(x)$.

Let us close this section by a short remark about the connection
between the Poisson-$\sigma$-model and a family of
quantum mechanical systems defined by the Poisson structure $P$ on $N$:
As $P$ is degenerate in general, $N$ is not the
phase space of a quantum mechanical system. But $P$ induces a
nondegenerate Poisson structure on each of the symplectic leaves,
whose inverse is the symplectic form $\Omega_S$. It is well known,
that a symplectic form on a manifold gives rise to a quantizable phase
space, iff it is integral (cf., e.g., \cite{Woodhouse}). (For $N=R^3$,
$\{X^i,X^j\}=\e_{ijk}X^k$ this expresses the fact that spins are
either integer or half integer).  Thus the integral symplectic
leaves are quantizable spaces. For trivial first homotopy of the
symplectic leaves we find (cf.\ Eq.\ (\ref{sphst})) that
the Hilbert space of the Poisson-$\s$-model contains exactly one state
for any quantizable quantum mechanical system on $N$.

Moreover, a natural action (corresponding to a vanishing Hamiltonian)
on these phase spaces is given by
the symplectic potential $\int d^{-1}\Omega$. So the phase factor in
the expression \rz wavfu for the wavefunction of the
Poisson-$\sigma$-model is precisely the exponential of the action of
the quantum mechanical system defined on the symplectic
leaf $S \subset N$ corresponding to the chosen value for the $X^I$.

\section{Outlook}

In the previous section our analysis was restricted to local
classical solutions and quantum states on the cylinder. It would be
desirable to generalize the results taking into account all
global aspects of the theory for arbitrary world sheet topology.

The calculation of the partition function on arbitrary Riemann
surfaces would be a significant step into
 this direction and a path
integral calculation seems to be a reasonable approach to this
problem. Quite recently progress has been made on this issue:
at least we understand now, how the restriction to the integral
symplectic leaves arises from the point of view of path integrals.
But still work has to be done to obtain the final result.
So let us close this talk by
shortly giving you some insight into this problem.

Our starting point is the functional integral
\be
  Z_g(f)=\int [dA][dX]...f\,\exp i\int(AdX+P\,A\wedge A +...)
\eel pi1
The dots indicate some additional fields and terms which will enter
due to gauge fixing and $f$ denotes an arbitrary $P$-invariant function
on the target space $N$, representing some observable.
If we now think of having Casimir-Darboux coordinates $\{X^I,X^\a\}$
on $N$, the gauge transformations take the simple form
$\d_\ep X^\a =  \ep_\a (x) P^{\a\b}$, $\d_\ep A_i = d\ep_i$
and allow to fix the gauge by gauge conditions on
$X^\alpha$ and $A_I$. Then integration over $A_\alpha$ is not affected
by the gauge fixing procedure and can be performed directly.
This leads to
\be
  Z_g(f)=\int [dX][dA_I]...f\,\exp i(A_IdX^I + \Omega_{\alpha\beta}
   dX^\alpha dX^\beta +...)
\eel pa2
where $\Omega$ is the symplectic form on the symplectic leaves.
With a Hodge decomposition $A_I=d\alpha_I+*d\beta_I+\gamma_I$ (with
respect to an arbitrary positive definite metric on the genus-g Riemann
surface $M_g$ playing the role of the world sheet) the gauge fixing
will not
affect $\beta_I$ and we can perform the integration over $\beta_I$. It
leads to a $\d$-function
yielding $X^I$ to be constant. So, denoting by $X^I_{(0)}$ the constant
mode of $X^I$, one finds
\be
   Z_g(f)=\int dX^I_{(0)}[dX^\alpha]...f\exp
      i\int_{Im(X^\alpha)}\Omega(X^I)+...
\eel pa3
Gauge fixing the $X^\alpha$ will reduce the integral over all maps
$X^\alpha :M_g \to S$ from the world sheet $M_g$ to a symplectic leaf
$L$ to a sum over homotopy classes of maps. So the partition
function takes the form
\be
 Z_g(f)=\int dX^I_{(0)}\mu(X^I)f(X^I)\sum_{[M_g\to S(X^I_{(0)})]}
   i\int\Omega(X^I)
\eel pi4
Here we plugged in some measure $\mu$ which may arise due to ghost
contributions. Now specializing to the case, where the first homotopy
class of the symplectic leaves $S$ is trivial, the homotopy class of a
map $M_g\to
S(X^I_{(0)})$ is characterized by winding numbers $n_a$ around
generators $\rho_a$ of the second homotopy of $S(X^I_{(0)})$. Then
\be
   \sum_{[M_g\to S(X^I_{(0)})]} \exp i\int\Omega(X^I) =
       \sum_{n_j\in Z}\exp i \sum_j n_j \int_{\rho_j}\Omega  =
       \sum_{m_i\in Z} \prod_i \delta (\int_{\rho_j}\!\!\!\Omega\,-m_i)
\eel pa5
Here Vol denotes the symplectic volume.
So we get the restriction to integral symplectic leaves.

Still the measure $\mu$ has to be determined. Unfortunately, we do not
know the general answer to this problem.

While the determination of the phase space of the theory and the
calculation of partition functions for world sheets of arbitrary
topology is most interesting from the point of view of topological
field theory, there is another open question, particularly
interesting from the point of view of quantum gravity, namely the
coupling of matter to Poisson-$\sigma$-models. Work should also be
done in this direction.

Hopefully we could convince the audience that Poisson-$\sigma$-models
are a most interesting subject deserving further investigation: They
involve a nice mathematical structure, give new insight into
theories already analyzed in the literature and provide us with new
examples of topological field theories.

\vspace{5ex}
{\Large\bf Acknowledgment}
\vspace{3ex}

We are deeply indebted to A.Alekseev for his contributions to our
collaboration.


\begin{thebibliography}{00}
    \bibitem{Blau} D. Birmingham, M. Blau, M. Rakowski and G. Thompson,
{\em Phys. Rep.} {\bf 209}, 129 (1991). E. Witten,
{\em J. Geom. Phys.} {\bf 9} (1992) 303.
   \bibitem{CGHS} C.G. Callan, S.B. Giddings, J.A. Harvey,
A. Strominger, {\it Phys. Rev.} D 45/4 (1992) R1005.
   \bibitem{gravmany}  D. Amati, S. Elitzur, E. Rabinovici, {\em On
Induced Gravity in 2-d Topological Theories}, preprint hep-th/9312003.
H.\ Verlinde, in {\em The Sixth
 Marcel Grossmann
Meeting on General Relativity}, edited by M.\ Sato (World Scientific,
Singapore, 1992).  D.\ Cangemi and R.\ Jackiw, {\em Ann. Phys.} (NY)
{\bf 225}, (1993) 229.
D. Christensen and R.B. Mann,
{Class. Quan. Grav.} {\bf 9} (1992) 1. M.O. Katanaev,  {\em J.
Math. Phys.} {\bf 34}/2 (1991) 700. H.\ Kawai and R.\ Nakayama,
{\em Quantum $R^2$ Gravity in Two Dimensions}, preprint KEK 92-212.
W.Kummer, D.Schwarz, {\em Phys. Rev.} D45 (1992) 3682. W.Kummer,
P.Widerin, {\em Mod. Phys. Lett.} A9 (1994) 1407.
  \bibitem{Rajetal} S.G. Rajeev, , Phys. Lett. B 212 (1988) 203.
 J.E. Hetrick and Y. Hosotani, Phys. Lett. B 230 (1989) 88.
E. Witten, {\em Commun. Math. Phys.} {\bf 141} (1991) 153.
S. Shabanov, {\em Phys. Lett.} {\bf 318B} (1993) 323.
\bibitem{GWZW} K.Gawedzki, K.Kupiainen, Nucl. Phys. {\bf B 320} (1989)
625. E. Witten, {\em Comm. Math. Phys.} {\bf 144} (1992) 189.
M. Blau and G. Thompson, Lectures on 2d Gauge Theories, hep-th/9310144.
  \bibitem{LNP} P.\ Schaller and T.\ Strobl, {\em Quantization of
Field Theories Generalizing Gravity-Yang-Mills Systems in Two Dimensions},
preprint TUW9402, to be published in LNP.
  \bibitem{p9}  P.\ Schaller and T.\ Strobl,
Poisson Structure Induced
(Topological) Field Theories in Two Dimensions, preprint TUW9403 and
hep-th/9405110,
to be published in {\em Mod. Phys. Lett. A}.
   \bibitem{WittenSigma} E. Witten, {\em Comm. Math. Phys.}
   {\bf 118} (1988) 411.
  \bibitem{Bauli} L.\ Baulieu and I.\ Singer,
{\it Comm.\ Math.\ Phys.} {\bf 125} (1989) 227.
  \bibitem{Migdal} A. Migdal,
{Sov. Phys. Jept.} {\em 42} (1976) 413.
  \bibitem{Kloesch}  T.Kloesch and T. Strobl,  in preparation.
   \bibitem{Kunst} D.\ Louis-Martinez, J.\ Gegenberger, G.\ Kunstatter,
{\em Phys. Lett. B} 321 (1994) 193. T.Strobl,
Dirac Quantization of Gravity-Yang-Mills
Systems in 1+1 Dimensions, preprint TUW9326 and hep-th/9403121, to be
publ. in Phys.Rev. D.
   \bibitem{CQG} P.\ Schaller and T.\ Strobl, {\em Class.Quan.Grav.}
{\bf 11} (1993)
   \bibitem{Semenov} M.A.Semenov-Tian-Shansky,
Dressing transformations and Poisson-Lie group
actions, In:
Publ. RIMS, Kyoto University {\bf 21}, no.6 (1985) 1237
   \bibitem{Anton} A.\ Alekseev, P.\ Schaller and T.\ Strobl, in preparation.
   \bibitem{Woodhouse} N.M.J.\ Woodhouse,  {\em Geometric
Quantization}, second Edition 1992, Clarendon Press, Oxford.
\end{thebibliography}
\end{document}